# Deep learning based surrogate model for first-principles global simulations of fusion plasmas

G. Dong [1+*], X. Wei [2+], J. Bao [3], G. Brochard [2], Z. Lin [2], W. Tang [1]

[1] Princeton Plasma Physics Laboratory, Princeton, NJ, U.S.A.
[2] University of California, Irvine, Irvine, CA, U.S.A
[3] Chinese Academy of Sciences, Beijing, China
[+] These authors contributed equally to this work
[*] gdong@princeton.edu

Abstract
The accurate identification and control of plasma instabilities is important for successful fusion experiments. First-principles simulations which can provide physics based instability information including the growth rate and mode structure are generally not fast enough for real-time applications. In this work, a deep-learning based surrogate model as an instability simulator has been developed and trained in a supervised manner with data from the gyrokinetic toroidal code (GTC) global electromagnetic simulations of the current driven kink instabilities in DIII-D plasmas. The inference time of the surrogate model of GTC (SGTC) is on the order of milliseconds, which fits the requirement of the DIII-D real-time plasma control system (PCS). SGTC demonstrates strong predictive capabilities for the kink mode instability properties including the growth rate and mode structure.

# I. Introduction

In real-time toroidal plasma experiments, accurate physics-based information of plasma instabilities can provide important guidance for successful plasma control [Strait 2019]. For example, the neoclassical tearing mode (NTM) [LaHaye 2006] is one of the most commonly observed causes for plasma major disruption [de Vries 2011], which is the abrupt loss of plasma confinement accompanied by large instantaneous energy transport that can damage the experimental device, especially for larger future tokamaks. The identification, predictions and control of the plasma perturbations that can excite NTM, such as the sawtooth oscillations [Sauter 2002, Chapman 2012], form the basis for effective NTM control and early disruption alarms. ITER is the next-step international toroidal fusion reactor towards unlimited carbon-free energy [ITER website]. Currently, the construction and design of ITER operations and the plasma control system (PCS) are partially dependent on the empirical extrapolation of instabilities, transport and other confinement properties from smaller experimental devices around the world [Hawryluk 2019, Strait 2019]. First-principles based study of plasma instabilities can improve the understanding and prediction of the dynamics and transport at the plasma core [Lin 2002, Jardin 2015] and plasma edge [Groebner 2013, Chang 2017] in future fusion devices such as ITER. Plasma instabilities such as magnetohydrodynamic (MHD) modes, micro-turbulence and alfven eigenmodes are widely studied with the utilization of various types of physics models and computational algorithms, including fluid MHD codes [Cheng 1992, Breslau 2009], gyrokinetic Eulerian codes [Kotschenreuther 1995, Holland 2012], gyrokinetic particle-in-cell codes [Lin 1998, Chang 2004] and etc. While these physics-based simulations can provide accurate growth rate, mode structure, and driving mechanisms for various plasma instabilities in their linear [Taimourzadeh 2019] and nonlinear stage [Wang 2005] for realistic experimental equilibrium, the first-principles simulations can be expensive computationally. Simulation time for physical instabilities using gyrokinetic particle-in-cell codes is often on the order of hours on modern GPUs [Madduri 2011], making the direct application of these codes in real-time experiments infeasible.

On the other hand, statistical methods including machine learning models have been applied in the PCS to predict plasma behaviors [Rae 2019, Fu 2020, Kates-Harbeck Nature 2019]. Recently deep learning based models have achieved promising results in disruption predictions [Kates-Harbeck Nature 2019] and the prediction of perturbed magnetic signals [Fu 2020, Lyons 2018]. Deep learning based models have also been applied for building emulators to aid first-principles simulations [Miller, 2020]. Here we present the first results on building a deep-learning based surrogate model as a physics-based instability simulator, trained based on data from global gyrokinetic toroidal code (GTC), which has performed thousands of electromagnetic simulations in the fluid limit by suppressing all kinetic effects and using the DIII-D experimental equilibria. As an initial work in this area, we trained the surrogate models on simulation results of the linear internal kink modes, which are commonly observed [Wong 2000, Menard 2005, Xu 2012] current driven MHD instabilities that are closely related to the sawtooth activities [Porcelli 1996], fishbone modes [Heidbrink 1990, F. Wang 2013], and NTMs [Chang 1995]. Compared with the hours-scale gyrokinetic simulation time of GTC with kinetic effects for the internal kink mode, the inference time of the surrogate model of GTC (SGTC) is on the order of milliseconds, which fits the requirement of the DIII-D real-time PCS. SGTC demonstrates strong predictive capabilities of the linear kink mode instability properties including the mode growth rate and mode structure. The output of SGTC contains physics-based instability properties that can complement experimental measurements and provide more targeted guidance to the PCS. SGTC can also serve as a physical simulator in other plasma predictive algorithms, enabling the development of future AI-based plasma control algorithms.

In the rest of this paper, we introduce the design and workflow of SGTC in section 2, followed by the data properties and GTC simulations of the kink instabilities in section 3, and finally we present the training details and predictive performance of SGTC models in section 4.

## II. Deep learning based surrogate model of GTC (SGTC)

The design workflow of SGTC is shown in Figure 1. As shown by the deep blue solid arrows in the left panel, experimentally measured signals including the zero-dimensional scalar signals such as plasma core density, one-dimensional profile signals such as the plasma density profile, and two-dimensional signals such as the magnetic field, are used as inputs to SGTC. The inputs can easily be extended to higher dimensional measurements in future work. SGTC then outputs the plasma instability information, such as the mode growth rate, frequency, and global mode structure as illustrated by the perturbation snapshot in DIII-D plasma poloidal plane shown in the lower left panel. The instability properties can in turn be used in the PCS as inputs to control algorithms, as shown by the red arrow. SGTC output can also be fed to plasma predictive models, which then output plasma state predictions to the PCS. SGTC architecture is shown in the green box in the right panel. The zero-dimensional features are input to Nf1 fully connected layers, which outputs the zero-dimensional features. The high dimensional inputs are fed into a set of Nc convolutional layers with dropout layers, and then Nf2 fully connected layers. This output as high dimensional features will be concatenated with the zero-dimensional features, and then goes through the fully connected output layers. For mode stability and growth rate prediction models, the output dimension is 1, and for poloidal mode structure prediction models, the output dimension is $M_\psi * M_\theta$, where $M_\psi$ is the radial grid number, and $M_\theta$ is the poloidal grid number. The hyperparameter tuning processes of this network will be introduced in section 4.

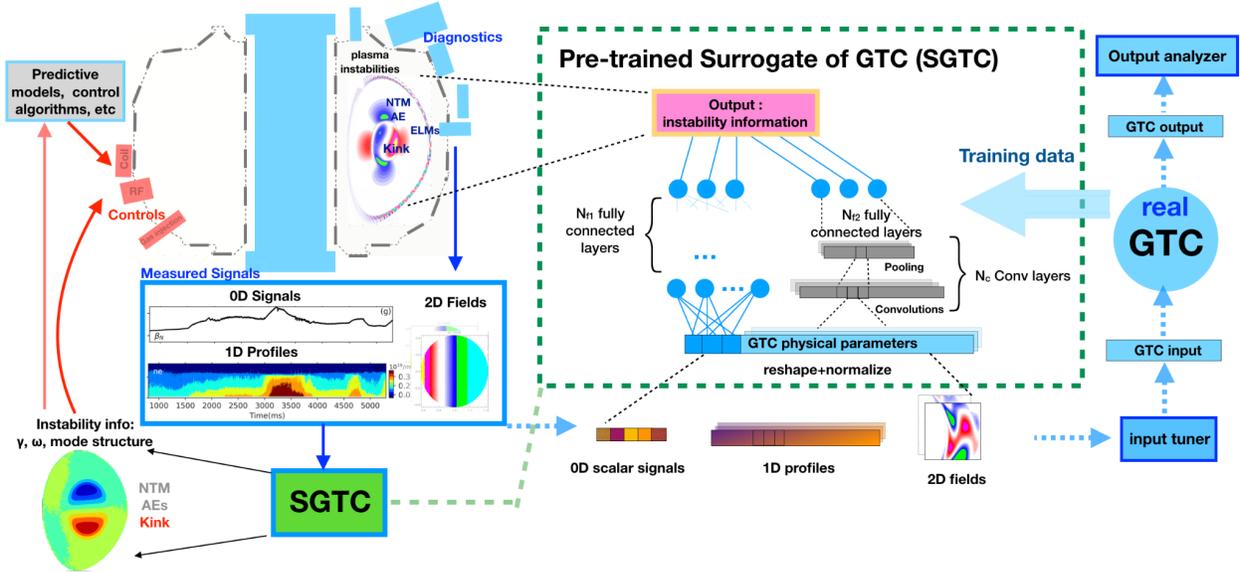

Figure 1. The workflow of SGTC. Measured signals are used as inputs to SGTC as shown by the deep blue solid arrows. SGTC then outputs the plasma instability information, such as the mode growth rate, frequency, and global mode structure as illustrated by the perturbation snapshot in DIII-D plasma poloidal plane. The instability properties can in turn be used in the PCS as inputs to control algorithms, as shown by the red arrow. SGTC output can also be fed to plasma predictive models, which then output plasma state predictions to the PCS. SGTC architecture is shown in the green box in the right panel. GTC data are generated and used for training, validation and testing of SGTC.

GTC first-principles simulation data are generated and used for training, validation and testing of SGTC. GTC is a global gyrokinetic simulation tool that has been developing since 1995 [Lin 1995], and has been validated for simulations of different types of plasma instabilities in DIII-D, JET, EAST, KSTAR, HL-2A tokamaks, W7-X and LHD stellarators, and C2 field-reversed configuration. The first-principles GTC simulations with associated theory and experimental observations have led to scientific discovery in turbulence self-regulation by zonal flows [Lin 1998], zonal flow damping [Lin 1999], neoclassical transport [Lin 1997], transport scaling [Lin 2002], wave-particle decorrelation [Lin 2007], energetic particle transport [Zhang 2008], electron transport [Xiao 2009], nonlinear dynamics of Alfven eigenmodes [Zhang 2012], localization of Alfven eigenmodes [Z. Wang 2013], driftwave stability [Schmitz 2016], transport bifurcation [Xie 2017] in fusion plasmas. Its simulation of current driven kink instability [McClenaghan 2014] has recently been benchmarked against different codes [Brochard 2021]. In this study, we utilize GTC to create a database of the internal kink instabilities in DIII-D plasmas.

As the first-step towards synthetic instability simulations, we run GTC electromagnetic linear simulations of the non-tearing n=1 instabilities in the ideal MHD limit with equilibrium current and compressible magnetic perturbations [Holod 2009, Dong 2017], where n is the toroidal mode number, for 5758 equilibriums selected from DIII-D experiments from magnetic EFIT01 [Lao 1985] and motional stark effect (MSE) EFIT02 [Lao 2004]. The complete GTC model formulation is described in detail in [Wei 2021]. After completing GTC runs, we then performed supervised training of SGTC models with these DIII-D equilibria and GTC output data. In most of the GTC simulations finding unstable modes, the mode structures resemble those of the internal kink modes with dominant m=1 component in electrostatic potential near the q = 1 rational surface [Rosenbluth 1973], where m is the poloidal mode number, and q is the plasma safety factor. In the following text, we refer to the n=1 MHD instabilities in GTC and SGTC outputs as the kink instabilities or the kink modes. SGTC models trained for these data can be considered as internal kink mode simulators for DIII-D plasmas.

## III. Data and GTC simulations

A list of DIII-D archived data that are used as GTC inputs are shown in table 1. Time sliced data are selected randomly from shot # 139520 to shot # 180844, with the condition that the listed data are available, and internal kink mode is possibly present at the time slice of interest. The presence of possible internal kink mode is determined from the combination of measured magnetic perturbations from the Mirnov coils and the minimum safety factor $q_{min}$ from EFIT [Lao 1985]. EFIT02 data usually provide finer magnetic pitch angle (i.e., safety factor q) information, as it includes input information from the MSE diagnostics [Wróblewski 1992]. Among the 5758 DIII-D equilibria we simulated for this work, 2872 are based on magnetic EFIT01 output, and 2886 are based on MSE EFIT02 output. These simulations have been carried out in 12 GTC runs, each simulating 500 DIII-D experiments in parallel using 2000 nodes of the Summit supercomputer at ORNL for about 30 minutes (Summit has about 4700 nodes).

| shot number | data |
|---|---|
| 5758 equilibriums from shot # 139520-180844 | EFIT01 gfile |
| | EFIT02 gfile |
| | ZIPFIT01 electron temperature profile |
| | ZIPFIT01 electron density profile |
| | ZIPFIT01 ion temperature profile |
| | Magnetic perturbation (mpi66M307D) |
| | Magnetic perturbation (mpi66M322D) |

Table1. List of DIII-D experimental data used as inputs for GTC simulations. Magnetic perturbation signals are only used for the selection of equilibriums.

As shown by the GTC data generation workflow on the right of the dashed green box in Figure 1, we first use an 'input tuner' to generate GTC inputs from experimental data. In this step we utilized the ORBIT code [White 1984] to convert equilibrium data to Boozer coordinates. With the GTC inputs, global electromagnetic simulations in the MHD limit are run for each input (i.e., each equilibrium of the DIII-D experiments) for 3000 time steps, with time step size $0.01\frac{R_0}{C_s}$, where $R_0$ is the major radius and $C_s = \sqrt{(T_e/m_i)}$ is the ion sound speed with $T_e$ as the on-axis electron temperature and $m_i$ as the ion mass. Typical total simulation time for each equilibrium is on the order of 0.1 ms for the physical duration of the experiments. After completing GTC simulations, we used an 'output analyzer' to examine the output data, exclude numerical instabilities, and prepare proper target data such as the mode growth rate, and poloidal eigenmode structure for SGTC. The n=1 mode growth rate is calculated with a linear fit of the perturbed parallel vector potential from the last 1000 time steps of the simulations.

Among the 5758 DIII-D equilibria that GTC simulated in this study, 1972 equilibria have unstable n=1 kink modes. GTC simulation results of all the 1972 unstable cases and 2531 stable cases are used for training (80% of the data), validation (10% of the data), and testing (10% of the data) of SGTC. The histogram of the unstable mode growth rate in training, validation and testing dataset for the neural networks are shown in Figure 2. Since the simulations are in the ideal MHD limit without kinetic effects, the linear growth rate can be large, for example above 400kHz. The distribution of the mode instability in physical parameter space is shown in Figure 3. The y axis in Figure 3 is defined as $\delta\beta_p = -\frac{R_0^2 \int_{}^{r_1} p' r^2 dr}{B_0^2 r_1^4}$, where $r_1$ denotes the minor radius of the q=1 surface, $p'$ denotes the radial derivative of the plasma pressure profile, and $B_0$ denotes the toroidal magnetic field. $\delta\beta_p$ is a relevant parameter for the kink instability in linear ideal MHD theory [Bussac 1975].

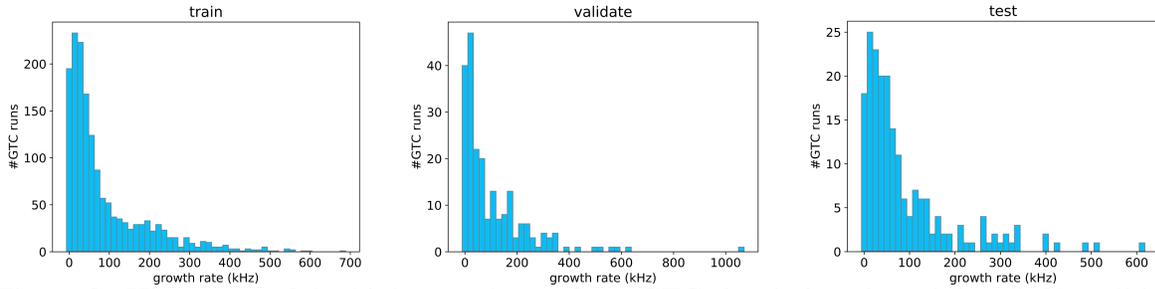

Figure 2. Histogram of the kink growth rate from GTC simulations in training (left), validation (middle) and test (right) dataset. Simulation results with stable internal kink mode are not plotted.

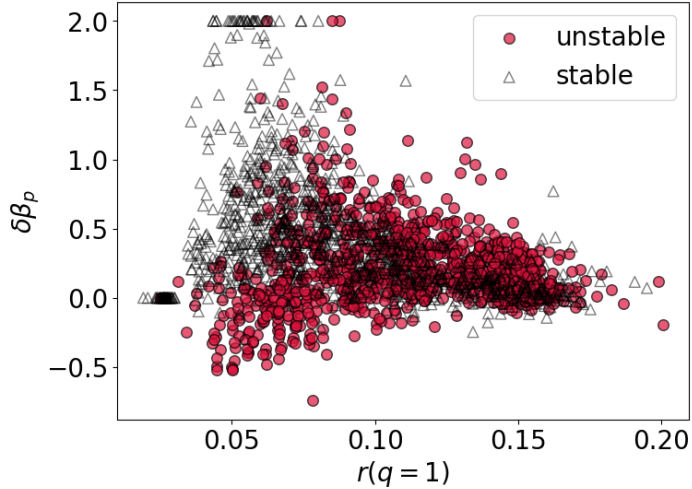

Figure 3: Kink instability distribution in parameter space from GTC simulations.

IV. Training and Performance of SGTC

In SGTC internal kink simulators, we trained neural networks to predict the mode instability (which is a binary value to indicate whether the equilibrium of interest is kink unstable), mode growth rate, and two dimensional poloidal mode structure. The first predictive task of SGTC is the prediction of mode instability. In addition to deep learning based models, we also tested the performance of classical algorithms in which models can not be easily extended in a stacked layers fashion. In Figure 4, we show comparison of the predictive capabilities of the neural networks and of the classical algorithms. Inputs of the models are described in section 2, and the output of the model is an instability score. A threshold for the instability score can be varied to achieve a receiver operating characteristic (ROC) curve for each model as shown in Figure 4, where the x axis is the true positive rate (TPR) and the y axis is the false positive rate (FPR). We used the area under the ROC curve (AUC) as the metric to evaluate model performance.

We trained multiple classical models in the sklearn package [sklearn], and the best performing one is the random forest model. We hand-tuned both the random forest model and the neural network for the instability prediction on the validation dataset, and reported the ROC curve on the test dataset. The AUC for the deep learning based model (DL) and the random forest (RF) model on the test dataset are 0.945 and 0.927 respectively. Figure 4 shows that the neural network outperforms the random forest model near the optimal threshold regime, where on the ROC curve the distance to the upper left corner (true positive rate = 1 and false positive rate =0) is minimized. At the optimal threshold based on the validation dataset, the prediction results on all test data from magnetic EFIT01 equilibria are visualized in Figure 5 in physical parameter space. The accuracy for the neural network and the random forest model on the test set is 0.89 and 0.85, respectively.

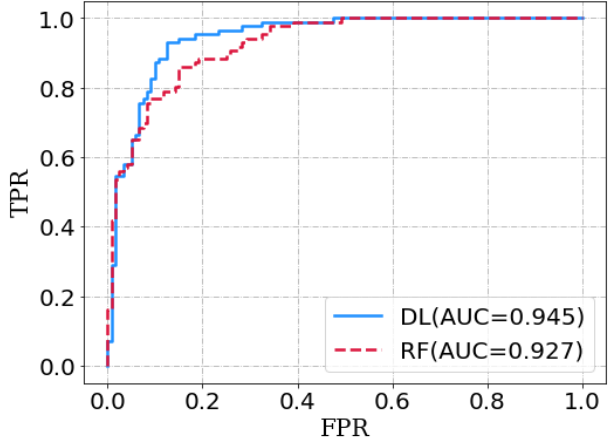

Figure 4: ROC curve of instability predictions for the neural network in a solid blue line and the random forest model in a dashed red line on the test set. The area under the ROC curve (AUC)s are 0.945 and 0.927 for the neural network and random forest respectively.

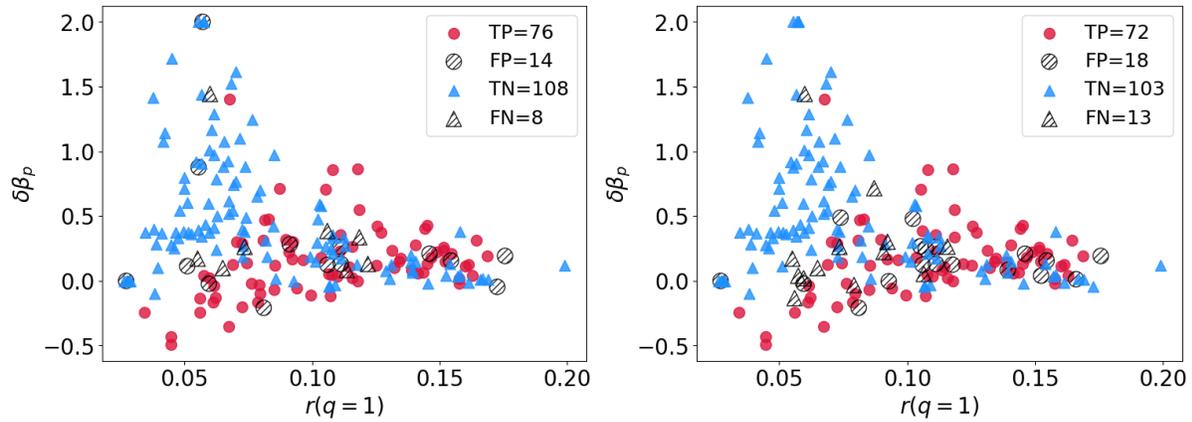

Figure 5: Prediction of kink instability on test dataset from deep learning method in the left panel and random forest method in the right panel. Solid red dot represents the true positive (TP), solid blue triangle represents the true negative (TN), shaded circles represent the false positive (FP), and shaded triangles represent the false negative (FN).

For the second task of SGTC predictions of the mode growth rate, we performed automatic hyperparameter tuning, by randomly generating 200 models in the hyperparameter space. The mean squared error is used as the loss function in this regression problem. After training all 100 models, we select 10 best performing models based on their validation loss, and report the ensemble result of these 10 models. A visualization of the test result is shown in Figure 6, where the left panel shows the true value of the growth rate vs the predicted value of the growth rate. The right panel of Figure 6 shows a histogram of prediction error. There are some significant under predictions for several test data points with the growth rate greater than 200 kHz. The accuracy of the prediction decreases when the growth rate becomes large, possibly due to the small number of training data in the large growth rate regime, as shown in the histogram of growth rate in Figure 2. In Figure 7, we compared SGTC performance with a simplified analytic formula of the internal kink instability [Rosenbluth 1973] $\gamma \propto k_z^3 V_A r^2 (q = 1)$, where $k_z$ denotes the parallel wave number, and $V_A$ denotes the Alfven velocity. In the left panel of Figure 7, we showed a comparison of the mean squared error of the prediction of the growth rate from random guess, the analytic formula and SGTC for all the data in the test set. The random guess is randomly generated values from uniform distribution ranging from 0-10kHz, where 10kHz is around the mean growth rate of the unstable cases. The error bar is the standard deviation divided by the square root of the number of test data points. We normalized the analytic formula such that the mean growth rate on the test set matches the true value (which can not be available in realistic scenarios). The high error level of the analytic formula from GTC results shows that GTC simulation of the kink mode in the MHD limit for realistic DIII-D geometry yields significantly different results from simplified analytic estimations. The mean squared error of SGTC predictions of the mode growth rate for the test dataset is 4.4e3 kHz$^2$, significantly lower than the analytic formula estimation. In the right panel of Figure 7, we compared the mean squared error of the prediction of the growth rate for unstable cases with growth rate smaller than 50 kHz. The yellow bar represents the difference between GTC simulation and simulation result from four other MHD codes M3D-C1, GAM-solver, NOVA, and XTOR-K [Brochard 2021] for the DIII-D shot number 141216 at 1750 ms, which has unstable kink mode in the MHD limit with a growth rate around 50 kHz. The details of this benchmark is presented in [Brochard 2021]. These results show that SGTC has strong predictive power for the kink linear growth rate. With more training data and advanced algorithms, its performance can be further improved to better represent GTC simulation results.

We would like to highlight that the average inference time of the ensemble model on NVIDIA V100 GPUs is 0.88 ms. With parallel algorithms, this inference model would fit the run time requirement of the DIII-D PCS. This facilitates the incorporation of the physics-based instability information from the first-principles based global electromagnetic simulations into the PCS of modern tokamaks.

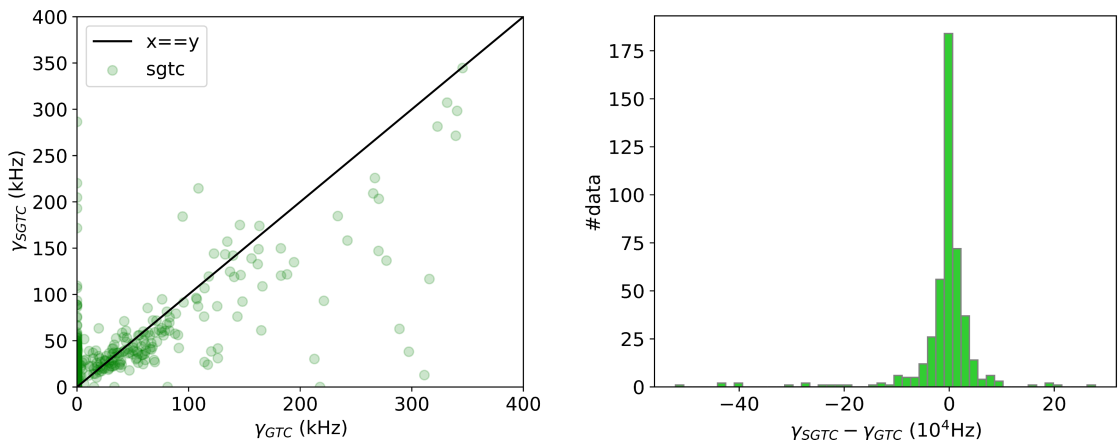

Figure 6. Prediction results of the kink growth rate for entire test dataset. The left panel visualizes the true value of the growth rate vs the predicted value of the growth rate. The solid

black line indicates x == y where perfect predictions occur. The right panel shows a histogram of prediction error.

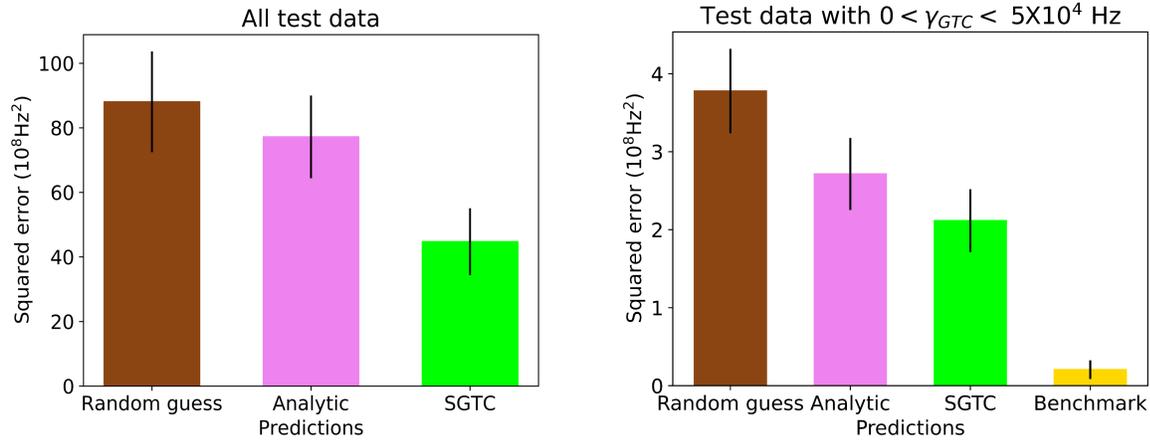

Figure 7. Comparison of SGTC performance for all test dataset with random guess and analytic formula in the left panel. Right panel shows the comparison of SGTC performance with random guess, analytic formula for test data with kink growth rate smaller than 50 kHz. Yellow bar represents the difference between GTC and four other MHD-based simulations [Brochard 2021] for DIII-D shot #141216 at 1750 ms.

For the third task of the SGTC predictions of the poloidal mode structure, we used the linear mode structure of the perturbed electrostatic potential and perturbed parallel vector potential from GTC simulations. Examples of the poloidal mode structures from GTC simulations are shown in the upper panels of Figure 8. We tuned the hyperparameter automatically by training 10 models in the random hyperparameter space, and selecting the model with the minimum validation loss. Visualization of the SGTC prediction of the mode structure is shown in the lower panels of Figure 8. A qualitative agreement is achieved for about 85% of the test dataset. An example where SGTC predictions of the mode structures agree well with the GTC output is demonstrated in the left panels of Figure 8. The right panels of Figure 8 show GTC and SGTC output of the poloidal mode structure for the equilibrium in shot #140510 at 3145 ms, where SGTC provides correct predictions for the mode structure in most parts of the simulation domain, but misses the fine structure near the magnetic axis. The SGTC prediction is actually a more "typical" internal kink mode structure compared to the ground truth. That means the model has learned the mode structure features statistically. In both cases, the model realizes the importance of q=1 surface to the mode structure.

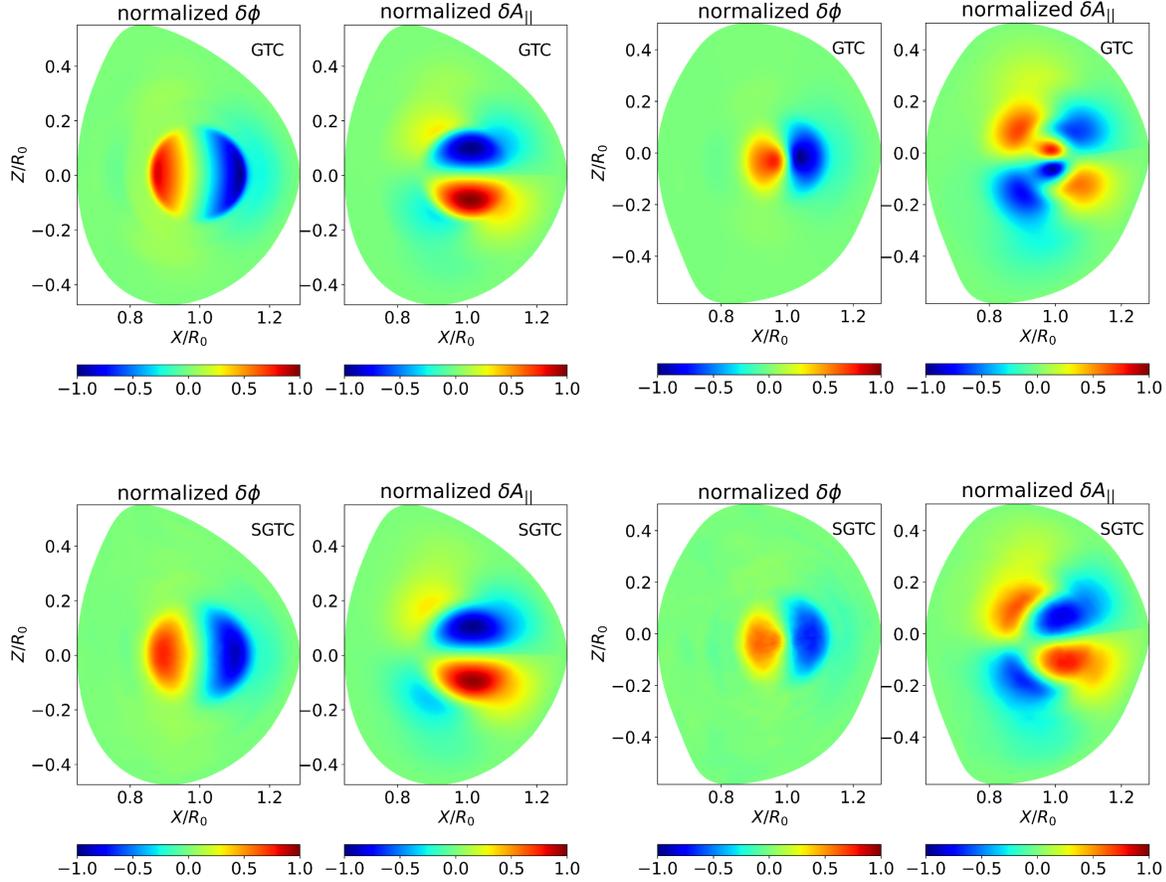

Figure 8. Example of SGTC prediction results of the mode structure of shot #162930 at 1820 ms using magnetic EFIT01 reconstruction and shot #140510 at 3145ms using MSE EFIT02 equilibrium reconstruction. Upper panels are GTC outputs and lower panels are SGTC outputs.

Finally, SGTC can be applied to simulate the entire duration of the DIII-D experiment for linear analysis of the internal kink instability. An example of SGTC 'simulation' of the internal kink mode evolution for shot # 141216 from 1500 ms to 5200 ms is shown in Figure 9. The left panels show the evolution of measured physical parameters, and the right panels show SGTC outputs. SGTC output shows that the n=1 kink mode is linearly unstable for most part of the shot, except at around 2-2.5s when the growth rate decreases. The mode structure shifts between q=1 and q=2 surfaces. This result can be qualitatively consistent with observations such as the mirnov coil data. Realistic equilibrium and instability scenarios are complicated by the fishbone modes, tearing modes, the kinetic effects and nonlinear dynamics of all the instabilities, which SGTC will be trained and tested on in future studies.

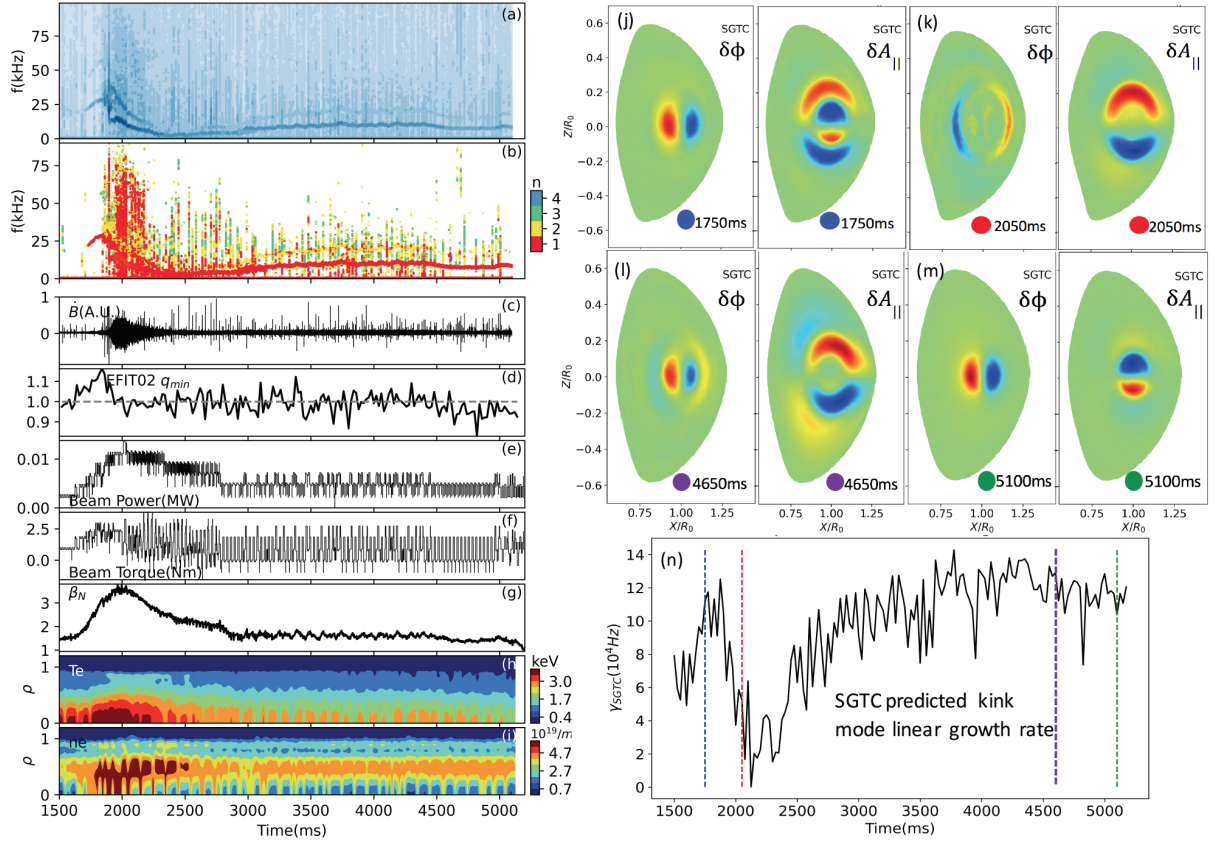

Fig 9. SGTC prediction of internal kink mode activity for shot #141216. Left panels show plasma signals from experimental measurements. Magnetic spectrograms from the Mirnov coils are shown in panels (a) and (b), where (b) shows the toroidal mode number n calculated from two coils. Panels (c)-(i) show magnetic perturbation amplitude, minimum safety factor from MSE EFIT02 equilibrium reconstruction, total input beam power, total input beam torque, normalized plasma beta, plasma temperature profile and plasma density profile respectively. The SGTC predicted time evolution of the internal kink mode growth rate is shown in panel (n), and the predicted mode structures for time slices 1750 ms, 2050 ms, 4650 ms and 5100 ms are shown in panels (j)-(m) respectively.

## V. Conclusions and future work

In this work we designed and constructed a framework for deep learning based surrogate models for first-principles global electromagnetic toroidal plasma simulations. As a demonstration of this new tool, we simulated 5758 DIII-D equilibria using GTC in the MHD limit. The surrogate models of GTC (SGTC) are trained to predict for the n=1 current driven internal kink instabilities in DIII-D plasmas. The SGTC internal kink simulators demonstrate strong predictive capabilities. SGTC shortens the simulation time by at least six orders of magnitude, and presents for the first time the possibility of bringing physics-based instability information from the first-principles based massively parallel simulations into the PCS of modern tokamaks.

This paper focuses on comparing SGTC output with GTC simulations results. Since the GTC simulation model used in this work is in the ideal MHD limit in the kink simulations, without kinetic effects, energetic particle effects, sheared flow, nonlinear dynamics etc, the training data may not be realistic in quantitative comparison with experimental data. This work demonstrates the verification of SGTC against GTC results for the internal kink modes, and the equilibrium database used in this work is based on magnetic EFIT01 and MSE EFIT02. For validation against experimental data in next-step work, kinetic equilibrium reconstructions [Lao 2004] will be considered. In the future, SGTC can be trained to output the instability properties of the fishbone, tearing mode, Alfven eigenmode, and microturbulence etc, along with their nonlinear dynamics and transport levels. The methodology of SGTC can also be applied to training emulators for other first-principles plasma simulations such as the MHD codes.

Acknowledgements

We would like to thank Dr. William Heidbrink, Dr. Brian Victor, Dr. Jayon Barr and other colleagues for their support and helpful discussions. GTC runs in this work are performed on Summit. Neural networks in SGTC framework are trained on Summit (https://www.olcf.ornl.gov/olcf-resources/compute-systems/summit/), tigergpu (https://researchcomputing.princeton.edu/systems/tiger) and traverse (https://researchcomputing.princeton.edu/systems/traverse).

This work is supported by the U.S. Department of Energy (DOE) SciDAC project ISEP and used resources of the Oak Ridge Leadership Computing Facility at Oak Ridge National Laboratory (DOE Contract No. DE-AC05-00OR22725) and the National Energy Research Scientific Computing Center (DOE Contract No. DE-AC02-05CH11231).

This work is partially based upon work using the DIII-D National Fusion Facility, a DOE Office of Science user facility, under Awards DE-FC02-04ER54698. This report was prepared as an account of work sponsored by an agency of the United States Government. Neither the United States Government nor any agency thereof, nor any of their employees, makes any warranty, express or implied, or assumes any legal liability or responsibility for the accuracy, completeness, or usefulness of any information, apparatus, product, or process disclosed, or represents that its use would not infringe privately owned rights. Reference herein to any specific commercial product, process, or service by trade name, trademark, manufacturer, or otherwise, does not necessarily constitute or imply its endorsement, recommendation, or favoring by the United States Government or any agency thereof. The views and opinions of authors expressed herein do not necessarily state or reflect those of the United States Government or any agency thereof.